# The Design and Investigation of the Self-Assembly of Dimers with two Nematic Phases

Z. Ahmed, C. Welch and G. H. Mehl*

Department of Chemistry, University of Hull, Hull, HU6 7RX, UK; g.h.mehl@hull.ac.uk

A series of non-symmetric dimers were synthesised containing either cyanobiphenyl or difluoroterphenyl moieties on one side and a range of long, short, bent, polar or apolar mesogens on the other side of the molecules. The dielectric anisotropy of the mesogens was varied systematically. The systems were characterised by differential scanning calorimetry (DSC), optical polarizing microscopy (OPM) and detailed X-ray diffraction (XRD) studies, both in the nematic and the $N_x$ phase. The results are compared and structure properties relationships are discussed. A model for the assembly in the $N_x$ phase is developed discussing $N_{tb}$ structures, coaxial helices, swiss roll structures and chiral domain formation.

**Introduction**

Since the report of an additional nematic phase, initially termed $N_x$, in cyanobiphenyl based dimers, research into this additional nematic phase has increased rapidly.[1,2] Other materials with nematic-nematic transitions have been investigated and sets of new materials have been reported.[2-4] This research has gained additional impetus as it was recognized from the beginning that in the low temperature nematic phase chiral domains are formed. Moreover very fast switching in these materials in the low microsecond regime with a linear electronic effect open up the potential for very fast displays, obviating the need for colour filters and being thus very energy efficient.[2,5] Surprisingly, the structure of this lower nematic phase is still being actively discussed.[2,5] Studies in thin films are consistent with a model where extremely short pitch helices (8-11 nm) are formed, with the pitch being not longer than two to three molecular lengths. These helices are almost two orders of magnitude shorter than those formed by cholesteric liquid crystals, with the mesogens twisting and bending, hence the term $N_{tb}$. This view is supported too by freeze fracture TEM studies,[7-9] though AFM results point towards the possibility of surface crystallisation.[6] Solid state NMR studies of samples in the bulk, used initially to support the simple $N_{tb}$ model[10] cannot, according to more recent results,[11] be employed for this without invoking very unusual molecular features, such as very fast migration along helix axes. The optical defect textures of the $N_x$ phase are distinctly different to that of a typical nematic and are more reminiscent of those of smectic phases.[12] The crucial experiment to distinguish between an $N_x$ and a layered phase are XRD studies, where the absence of intensities associated with layering confirms the nematic character. Theoretical interest has associated this low temperature nematic phase quickly with the potential splay bend and twist bend structures alluded to earlier[13] and developed in more detail theoretically as well as in simulation studies.[14] It is noted that a number of additional different varying models are currently discussed theoretically, ranging from double wall assemblies[15] to hexatic structures,[16] and other complex modes of assembly such as chiral domain structures,[11] – conceptually close to the random domain phase discussed for bent core molecules[19] are considered. The formation of bent conformers has linked this low temperature phase to research on bent–core systems which have an extremely rich phase morphology.

We note that there are nematic related phase structures with a $C_S$ symmetry and tetrahedric symmetry and polar structures that have also been proposed.[17-19] In the absence of full consistency of experimental results with one theoretical model the term $N_x$ is retained in this contribution. In this contribution we report the synthesis and investigation of the liquid crystal properties by OPM, DSC and XRD studies of a series non-symmetric dimeric systems and propose, directed by the experimental data, an assembly mode. The purpose of this study is to understand structure-property correlations of dimeric materials and the structure of the $N_x$ phase with the view to open this class of materials up to technological applications. Having in view the importance of dipoles for any LC device, these were systemticall vaired in the molecules. A summary of the synthesis of the materials is shown in Scheme 1. The materials are based on a nonyl spacer linking two mesogenic groups and were designed to exhibit a low temperature $N_x$ mesophase. This was to facilitate ease of investigation and the systematic exploration of the modulation of molecular structure on liquid crystal phase formation. For mesogenic groups the classical cyanobiphenyl group and the difluoroterphenyl system reported before are used. The molecule based on the mixed cyanobiphenyl /difluoroterphenyl system has been reported before[20] and the properties are listed here for clarity. Whilst either the cyanobiphenyl or the difluoroterphenyl group are kept constant on one side of the dimer, the length of the mesogenic group on the other was varied systematically. This was done in such a manner that the mesogenic group consisting of two aromatic rings contain either lateral fluorine or hydrogen atoms and the termini are formed by either pentyl or butyloxy groups, selected for similarity in their overall length. In order to explore the effects of molecular bend a nitrile functionalized thiophene unit was included in the series. The asymmetric final compounds **1, 2, 3** and **4** were prepared by two successive Suzuki coupling reactions. The first reaction was carried out with 4-cyanophenyl boronic acid with a large excess (4x) of the spacer **I1**. The spacer **I1** was synthesized as described previously.[21] The mono substituted compound was the major product with the conditions mentioned in the procedure, and a 10 times dilution was also tried but no significant difference was observed. The yield was observed to be 65% and the unreacted spacer was easily isolated by chromatography and reused. The second reaction was carried out with the appropriate boronic acid in excess this time to achieve complete conversion. The yields were found to be between 42-71% with the lowest yielding reaction containing cyanothiophene boronic acid.

This is possibly due to the deboronation of the boronic acid. Strongly electron withdrawing groups are known to promote deboronation in basic conditions[22]. The asymmetric final compounds **5, 6, 7** and **8** were also prepared by two successive Suzuki coupling reactions. The first reaction was carried out with difluorobiphenylpentyl boronic acid with a large excess of the spacer **I1** (4x). The yield was 65% and again the unreacted spacer was isolated by chromatography and reused. The second reaction was carried out with the appropriate boronic acid in excess this time to obtain complete conversion. The yields were found to be between 44-77% with the lowest yielding reaction again containing cyanothiophene boronic acid. For the details of the synthesis and the chemical characterisation see the ESI.

**Results and discussion**

All of the compounds show nematic phase behaviour. A summary of the results, based on the OPM and DSC investigations is listed in Table 1. For the materials bearing a cyanobiphenyl unit on one side containing thus in total four aromatic rings, a clear trend is detectable. The nonlinear mesogen **1**, closest in structure to the cyanobiphenyl analogue CB-C9-CB shows the highest LC phase stability, becoming isotropic at 101.1˚C a reduction of 23.1 ˚C when compared to CB-C9-CB. $N_x$ phase formation is destabilized too, resulting in a monotropic phase at 89.5 ˚C, compared to an enantiotropic $N_x$ phase at 107.3 ˚C for CB-C9-CB. This is attributed to the reduced packing efficiency of this non-linear molecule when compared to the parent system. For the linear mesogens the trend is as anticipated, and compound **3** without the lateral fluoro groups has the highest transitions of the series, forming a nematic phase at 81.9 ˚C. Introduction of two lateral fluoro groups in **2** decreases nematic stability to 43.8 ˚C and the $N_x$ phase becomes monotropic at 38.3˚C, about two degrees below the melting point of 40.6 ˚C, although it remains stable for several days at room temperature. Introduction of an ether function enhances phase stability with the material **4** becoming isotropic at 62.7 ˚C, though the $N_x$ phase is only monotropic at 54.9˚C, 1.4˚C below the melting point at 56.3˚C. The overall trend seen is that a reduction of molecular symmetry lowers the transition temperatures and the melting points with a concomitant reduction in the stability of the low temperature nematic phase. The transition enthalpies range from 0.46 kJ mol$^{-1}$ to 1.31 kJ mol$^{-1}$ and are in line with typical nematic materials, but noticeable is the much higher value of 2.10 kJ mol$^{-1}$ for CB-C9-CB. For the materials bearing a terphenyl unit at one end of the molecule the phase behaviour is considerably more complex. All these materials are non-symmetric with regards to the length of the aromatic groups, with the systems each bearing five aromatic rings in total. For molecule 5 with a thiophene 5-cyano group at the end a clearing temperature of 114.7 ˚C from the nematic is observed with only a monotropic $N_x$ phases detectable at 83.9 ˚C. For the linear molecules the differences are quite considerable and these values are substantially lower than those for the already reported molecules DTC5C9 and DTC9-CB, where respective isotropisation temperatures of 144.1 ˚C and 165.3 ˚C and $N_x$ phase formation at 109.8 ˚C and 120.3˚C for DTC5C9 and DTC9-CB have been reported.[7,20] Compound **6** with two lateral fluoro groups on the short mesogen melt**s** at 47.4˚C, and shows a highly ordered smectic phase up to 62.2 ˚C, where the material transforms to a SmA phase which becomes nematic at 71.6˚C before the materials turns isotropic 81.3 °C. Removal of the fluoro groups on one side in compound **7**, enhanced the phase stability and lead to an $N_x$ phase ranging from 78.2°C to 86.4°C with the material turning isotropic at 105.5 °C. Introducing a terminal butyloxy group changes the phase behaviour considerably as seen for compound **8**. A monotropic highly ordered LC phase is formed on cooling at 61.7 °C, with the melting point at 69.5 °C, when an $N_x$ phase is formed which is stable up to 82.8 °C, followed by a nematic phase before the material turns isotropic at 103.4 °C.

It is noted that the molar transition enthalpies for the lower temperature nematic phase for this set of molecules, ranging from 0.078 kJ mol$^{-1}$ for **5** to 0.37 kJ mol$^{-1}$ for **7** are higher than for the previous set of materials, however being still in the range of typical nematics.[22] It is interesting that the intermediate **I3** (see ESI) shows nematic phase behaviour up to 85 °C. A low temperature phase which was initially identified as $N_x$ based on optical studies, is verified to a SmA phase based on XRD investigations, for details see the ESI. In the process of changing the molecular structures, the dielectric anisotropies have been varied, lateral and terminal dipoles were introduced. The effect on the transitions are as expected. Lateral difluorogroups in the aromatic core reduce the transition temperatures, the introduction of terminal cyano groups and the introduction of ether linkages to the terminal chains enhance transition temperatures, when compared to methylene linked systems. The introduction of a bend in the mesogen, such as in compounds **1** and **5**, does not stabilize $N_x$ phase formation, however as monotropic phase behaviour was recorded close to the melting points, it can be assumed that small variations of the architecture could lead to enantiotropic $N_x$ phase formation. There is no indication that the position of dipoles or their symmetry plays a crucial role for $N_x$ phase formation, but they are important for dialling in the temperature range of $N_x$ phase formation.

**Polarising optical microscopy**

The optical defect textures of the $N_x$ phase are distinctly different to that of a typical nematic material. Though dark homeotropic areas can be observed, more typical are the observations of "tile" or "rope" like features often formed when cooling from the nematic. These defect textures can easily be mistaken for that of a smectic phase; indeed smectic focal conic like features are also found using three photon correlation spectroscopy.[12] Figure 1 shows the change of texture at 101 °C for compound **7** going from a nematic to Figure 1(b), containing rope like features and tile like features in the $N_x$ phase at 85 °C.

Figure 2(a) shows the plated texture close to the nematic-$N_x$ transition whereas Figure 2(b) shows the texture 2 °C lower at 33 °C. The texture observed can be described as polygonal with almost fish scale like resemblance. It is important to note that this texture change does not register as a thermal event in calorimetric studies. As both types of textures are kinetically stable it is likely the plated texture retains some characteristics of the preceding conventional nematic phase. If that were to be the case the natural texture of the $N_x$ phase would be the polygonal fish scale texture on untreated slides. This texture is shown below in Figure 3(a). A characteristic observation for the nematic phase is spontaneous homeotropic alignment under shearing. This phenomenon has been attributed to the small elastic constants of the nematic phase.[24] The same is observed for the $N_x$ phase and is shown in the shearing experiments in Figure 3(b). The dependence of the periodicity of the stripes to the thickness of the sample can be observed for the micrographs shown in Figure 4 and is in line with earlier observations.

**X-ray diffraction investigations**

The XRD investigation of samples in capillaries in the presence of a magnet field of ~0.5T perpendicular to the XRD beam were performed on heating and cooling. A number of general trends could be observed. For the wide angle reflections the difference between the nematic and $N_x$ phase is very small, the values for lateral distances of the molecules vary between 4.5 - 4.8 Å.

In the $N_x$ phase the wide angle maxima becomes sharper, suggesting a more ordered lateral register, resulting in a more defined positional ordering. Common for all materials is an excellent macroscopic ordering parallel to the external magnetic field in the nematic (N) phase, evidenced by equatorial wide angle maxima. At the transition to the $N_x$ phase this ordering is reduced. The maxima of the wide angle arcs are less well centred on the equator and more widely distributed, indicative of either loss of macroscopic order or some form of loss of long range orientational ordering. This is conceivable in a process where two mesogens linked at a fixed angle compete for orientation in an external field. Alternatively this could be explained by an increased mosaicity of the sample, or in other words domains are formed and the order within the domains can increase. This picture would tie in with the sharpening of the wide angle intensities, the observed macroscopic ordering by the external magnetic field is however lower than in the nematic phase. For the cyanobiphenyl containing systems the small angle intensities are very small, with intensity values considerably lower than those of 5CB, examined under similar conditions for which data was collected to be used as a reference (see Table 2). We note that the data collected here for 5CB is in line with results reported earlier.[25] Figure 5 shows diffractograms of **4** in the nematic phase at 58˚C and at 52˚C in the $N_x$ phase, with a vertical magnetic field. The loss of macroscopic ordering going to the $N_x$ phase is easily visually detectable, with the wide angle arcs less centred on the equator and more spread out. The very low intensity for the weak small angle reflections, hardly detectable for both the N and $N_x$, phase is noticeable and more clearly depicted in a pseudo 1D plot, shown in Figure 6 where the intensity of collected data against the diffraction angle 2θ is plotted.

For the materials containing a difluoroterphenyl group in the mesogenic unit the overall behaviour is very similar, but the stronger intensity of the small angle reflections is noticeable. The results for the investigated materials are summarized in Table 2. The maxima for the wide angle reflections shift for most of the investigated systems by 0.1 Å by going from the N to the $N_x$ phase, typically from 4.7 to 4.6 A, although there are exceptions such as for compound **5**, where there is no change and an overall lateral distance of 4.8 Å is measured. Though not unusual, these distances are slightly larger than for most nematics, where values of 4.5-4.6 Å are common. These results might simply be due to packing inefficiencies of the odd membered hydrocarbon structures, connected to the aromatic rings via a fairly rigid methylene group.

The apparent pseudo d-spacings derived from the weak small angle intensities are all quite similar in value, ranging between 15.9 Å – 21 Å, larger than a single aromatic mesogenic group, but slightly smaller than that of half of the full molecules, indicative of a local interdigitated structure of the $SmA_d$ type. The exception is molecule **1,** where the small angle intensities were so weak (see ESI for diffractogram) that is was not possible to define a pseudo d-spacing for the nematic phase. Looking at the temperature dependence of the small angle data it is noticeable that there is no clear trend observable for the investigated set of materials. For some of the materials on cooling to the $N_x$ phase from the nematic, a contraction in the value of the pseudo d-spacings is noted such as in material **2**, where a contraction from 17.5 Å at 40 °C to 15.9 Å at 35 °C was recorded. In compound **4** the contraction was just from 17.6 Å to 17.4 Å going from 56 °C to 42 °C and **5**, where the contraction was from 21.3 Å to 20.8 Å going from 90°C to 80 °C. For material **7** no detectable change in the apparent layer spacing of 19.2 Å could be detected. To quantify the low intensities in the small angle region further a simple ratio between the individual small angle and wide angle reflections is calculated termed $I_{SAX}/I_{WAX}$, using the integrated data assembled for the pseudo 1D plots. This indicates for the cyanobiphenyl based systems a value below 1, for most systems a value just below or above unity is found. Notable is that the cyanobiphenyl based systems tend to have values lower than 1. On the whole all these values are similar to those of which can detected for 5CB, the archetypal nematic material, for which data was collected in the nematic phase at 22.0 °C under the same conditions as for the dimeric materials, confirming the essentially low ordered character of the $N_x$ phase. Using the small angle intensities, it is possible to estimate the longitudinal correlation lengths. The calculated correlation lengths are a quantity used to indicate the length scale over which particle-particle correlations are lost and has been adopted to describe short range order. Correlation lengths were calculated using $\xi = c/\Delta q$ where *c* is the function used to describe the scattered intensity and $\Delta q$ is the full width at half maximum (FWHM).[26] For compound **1** and the nematic phase of compound **6**, the scattering intensity is described by a non-regular line shape and an accurate measurement of FWHM is not possible. The values for the correlation lengths are on the order of one mesogenic length, between 10-15 Å (see Table 2), and broadly similar to those found for 5CB[27-29] indicating that the materials are only really oriented over a very short range. The small changes in values for the correlation lengths, going from the nematic to the $N_x$ phase by only 1-3 Å, if at all are qualitatively indicative that no substantive increase in long range order takes place and the ordering is very short range indeed, even if one considers, as has been done previously,[30-32] that measured correlation lengths might underestimate true ordering by a factor of three or four. We note here that when a material shows a SmA phase such as observed for compound **6** at 50 °C, typical correlation lengths of 1176 Å, correlating to ~30 molecular lengths are detectable. A higher ordered low temperature phase of **6** termed SmX needs to be investigated further. Moreover if one considers that the molecules with polar end groups such as materials **1-4** exist in associated and non-associated species as has been explored in detail for 5CB[26] and for asymmetric dimers[31] changes in the organisation can be examined by monitoringchanges in diffraction intensities at various temperatures, we can conclude that surprisingly few changes take place, going from the nematic to the $N_x$ phase. Noticeable too is that none of the materials show cybotactic clustering of the SmC type in the nematic phase or in the $N_x$ phase. This is a feature that would be anticipated in tilted molecules and is a feature common in bent core systems.[31-33] Moreover, there is no indication of a diffraction pattern indicative of short pitch helical structures. An "X-shaped pattern" would be anticipated with the first meridional reflection indicative of the subunit axial translation of the molecules and off-meridonal intensities relating to the pitch of the helix. Patterns of this type have been observed for many natural and synthetic helical materials, with DNA being the most notable.[35] We note too, that when tight pitch chiral assemblies with some layered ordering are formed, as in the B4 phase or in LC nanofilaments, they are clearly detectable by XRD experiments as well as in TGB or smectic blue phases.[36-40] Given that the materials investigated here are low ordered one would have anticipated at least intensities in the form of a "straight line" in the 2D data of the small angle intensities of the oriented samples, provided a simple helix would be present. This was not found, but crescent like small angle intensities were detected with the maxima of the small angle intensities being at an angle of 90° to the maxima of the wide angle data. Occasionally on the transition from the nematic to the $N_x$ phase a small misalignment to the external magnetic field can be detected by ~5-6°, which suggests that even in capillaries with a diameter of 0.8 mm surface effects can play a role. The results require consideration of the possible assembly structure of the $N_x$ phase.

Based on XRD data we arrive at picture that on cooling from the nematic phase, the lateral packing tends to be closer, however there is no or very little increase in long range orientational ordering and macroscopic ordering decreases. This is consistent with results from calorimetric studies, which show that the transition enthalpies are very low, and the specific heat decreases when adding a low molar mass compound, such as 5CB, however $N_x$ phase formation still occurs.[20] Data from OPM experiments and shearing experiments indicates an increased viscosity on shearing going to the $N_x$ phase and textures which point to a highly ordered smectic like phase. The available data constrains model formation for the structure of the $N_x$ phase considerably. There is no positive experimental evidence based on XRD data gained in this study to support a nematic phase formed from helices with constant pitch and radii, and to the best knowledge of the authors, no experimental evidence has so far been reported in the literature on other dimers either. Short range ordering and closer lateral packing than that in the nematic phase is supported by experimental evidence. NMR data for samples in the bulk in other studies shows clearly the formation of chiral structures of opposite handedness and formation of domains.[9-11] On the other hand, thin film studies are clearly indicative of heliconical ordering, though recent studies suggest that splay components could play a role too.[41, 45] Closer inspection of the TEM reports show that the periodicities observed are not fully fixed in the 8-10 nm range, but shorter and longer pitches are found too, as well as domain structures.[8] This is quite different to cholesterics where, at a given temperature, only one defined pitch is observable. This is due to the different origin of chirality. For the dimers in the $N_x$ phase the dynamic interconversion of chiral conformations is slowed down. This occurs to such an extent, that when compared to the N phase, they persist sufficiently long, so that neighbouring molecules in the close vicinity arrange accordingly. A structural model needs to incorporate these findings and in the following an effort is made to reconcile the apparently contradictory findings using different experimental techniques. Fig 7a shows a schematic of the $N_{tb}$ model which is supported by thin film studies, with a uniform helical pitch of a few dimers in length. Fig 7b shows a variation of this structure, in the centre the tight pitch $N_{tb}$ structure is shown, which is surrounded idealized in a coaxial manner with assemblies of varying pitch, formed by molecules packing onto the tightly folded $N_{tb}$ like structure. The formation of such an assembly mode could explain the varying helix lengths observed by TEM experiments, as well the very small correlation lengths for the materials in XRD experiments. Moreover, for biomaterials such as cellulose this varying pitch mode with spiralling helicity explains the lack of the typical helical diffraction pattern for fibrous structures.[42,43] In this model the uniform ordering in the direction of the z-axis (helix axis) has been reduced, however additional ordering perpendicular to the main axis due to the surrounding enveloping spiralling helices is required. Incidentally this concept bears on first sight similarity with the double twist model for high chirality cholesterics.[43] Figure 7c represents a schematic showing how this level of ordering can be relaxed. The coaxial structure is reduced, the strict layering is broken and the helical pitch changes, shown in an idealized form spiralling out from the centre to the periphery in a type of swiss roll structure, a feature observed in some nanorods48 Nematic splay components which have been discussed in terms of the $N_{tb}$ phase could possibly be accommodated by such a model.[44-45] Figure 7d shows a further level of disordering as the length of the helix is now finite and the material is low ordered resulting in a domain structure, or to use polymer nomenclature in analogy, the persistence length of ordering in the z-axis is not large. Surprisingly, this depiction is close to the picture often used intuitively for chiral structures, such as gastropods or chiral swirls.[47-49] The difference between figs. 7c and 7d is that main axis is of finite length and pitch and radius taper, but they could interconvert, simply by the alignment of domains. Considering the very low correlation length found by XRD studies in the dimeric materials, structures such Figures 7a and 7b seem to be less likely than Figures 7c or 7d, which additionally could easily explain the formation of domains observed by NMR studies and by the XRD data. External fields could favour the structure shown in Figure 7b or Figure 7a, the currently discussed $N_{tb}$ structure. As the driving force for this assembly is just the self-sorting of enantiomorphic conformers[50,51] in the confinement of thin electro-optical cells with alignment layers and strong fields the formation of helices could be rationalized. This might, under the right conditions be close to the structure formed in single crystals[51] shown for CB-C7-CB to be a helical structure in the 8-10 nm range. This easy deformation of the domains by external fields would explain the staggeringly diverse range of defect textures formed by materials in the $N_x$ phase, where cell thickness, alignment layers and external fields play a crucial role. [1-8, 10, 12, 41-43] The easy change from a ground state could possibly explain too, the diverse AFM and TEM data. Such chameleon like assembly behaviour has not been discussed much for smectics, with the exception of a model for nematic biaxiality,[53] however it is more common for dimeric and oligomeric smectic systems.[54,55] The concept discussed here is akin to those developed to account for the formation of cybotactic clustering in essentially nematic materials.[51] There, on decreasing temperature smectic fluctuations occur and this results in the dynamic formation of species of essentially nanometric size. Cybotactic ordering has been observed for a wide range of materials ranging from bent core systems to nematic dimers, oligomers, polymers and dendrimers and often persists over the full temperature range of the nematic phase.[55] In these systems analysis is eased, as dynamic local layer formation can be interrogated by diffraction experiments, due to modulation of the electron density in the domains. In the case of the $N_x$ phase the fluctuations are not smectic like, they are chiral, due to the self-sorting of chiral conformers into dynamically assembling and disassembling conformers and domains and the ordering is still nematic; hence the analysis is more difficult. Finally, whether tightly packed chiral domains assemble in hierarchical superstructures, which may be responsible for the smectic like OPM textures as it has been discussed for short pitch smectics such as the

SmQ phase or for blue phases needs to be explored and concomitant technological exploitation following the understanding of the phase structure is to be expected.[38,39,45] Hence there is clearly a need to correlate the multitude of theoretical efforts with experimental data and to explore whether there are a number of nematic sub-phases and how they connect to low temperature highly ordered LC structures, such as the smectic (Sm) and bent-core (B) phases.

**Conclusions**

For the investigated systems the variation of the molecular symmetry by using mesogens of different length and shape and of different dielectric anisotropy, by introducing strong terminal or lateral dipoles or both was investigated. As a general trend it was found that the reduction of molecular symmetry reduces melting points and transition temperatures up to almost room temperature range. Shorter mesogens tend to have lower transition temperatures. The position of dipoles is whether positioned terminally or laterally is not critical for the formation of the $N_x$ phase as such, but they are of course critical for technological applications. Detailed XRD investigations show that all of the materials are low ordered in the nematic phase, similar to 5CB and that the orientational ordering, as defined by correlations lengths increases if at all only very slightly in the $N_x$ phase, though macroscopic loss or order and closer lateral packing was observed for most systems. Based on the available data, parameters for model formation are discussed, having in view the need to reconcile the results from complimentary experimental methods. Here the advantages of the $N_{tb}$ structure, coaxial helices, swiss roll structures and chiral domain formation in their interconversion are discussed.

**Acknowledgements**

The authors thank the colleagues in the EU project BIND for discussion and acknowledge the EPSRC projects EP/J004480/1 and EP/M015726/1 and the EPSRC NMSF in Swansea is acknowledged for mass spectrometry data.

**Notes and references**

[a]Department of Chemistry, University of Hull, HU6 7RX, United Kingdom; g.h.mehl@hull.ac.uk.

† Footnotes should appear here. These might include comments relevant to but not central to the matter under discussion, limited experimental and spectral data, and crystallographic data.

Electronic Supplementary Information (ESI) available: [details of any supplementary information available should be included here]. See DOI: 10.1039/b000000x/

SCHEMES, FIGURES and TABLES

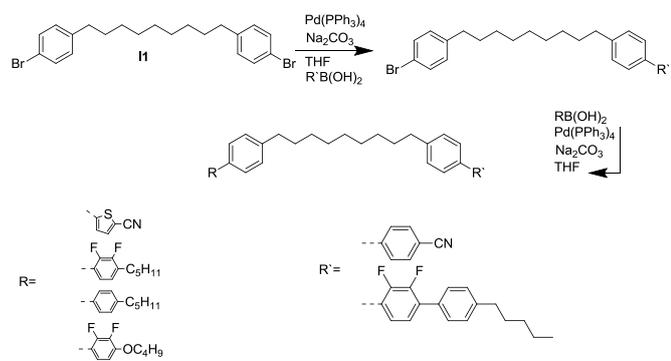

**Scheme 1: Synthetic scheme leading to the target materials**

| Table 1 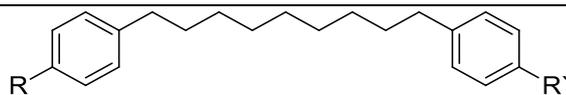 | | | |
|---|---|---|---|
| No. | R | R' | Transition Temperature [°C] $\Delta H$ [kJ·mol$^{-1}$] |
| CB-C9-CB[a] | -⌬-CN | -⌬-CN | Cr- 85.2- $N_x$ – 107.3 – N – 124.2 – Iso  0.86   2.10 |
| 1 | thiophene-CN | -⌬-CN | Cr – ($N_x$ 89.5) 90.4 – N – 101.1 - Iso  0.83   0.10 |
| 2 | -⌬(F,F)-C$_5$H$_{11}$ | -⌬-CN | Cr – ($N_x$ 38.3) 40.6 – N – 43.8 – Iso  1.31   0.078 |
| 3 | -⌬-C$_5$H$_{11}$ | -⌬-CN | Cr – 81.9 – N – 85.8 – Iso  0.37 |
| 4 | -⌬(F,F)-OC$_4$H$_9$ | -⌬-CN | Cr – ($N_x$ 54.9) 56.3 – N – 62.7 – Iso  0.46   0.15 |
| 5 | thiophene-CN | -⌬(F,F)-⌬-C$_5$H$_{11}$ | Cr – ($N_x$ 83.9) 89.7 – N – 114.7 – Iso  0.078   0.52 |
| 6 | 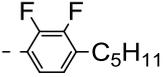 | 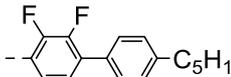 | Cr – 47.4 –SmX – 62.2 – SmA – 71.6 – N – 81.3 – Iso<br>3.98          0.98 |
| 7 | 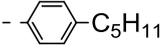 | 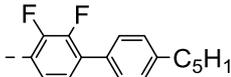 | Cr- 78.2- $N_x$ – 86.4 – N – 105.5 – Iso<br>0.37          0.99 |
| 8 | 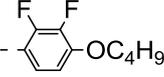 | 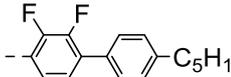 | Cr – (SmX 61.7) 69.5 – $N_x$ – 82.8 – N – 103.4 – Iso<br>1.01     0.10     0.38 |

Transition temperatures and enthalpy vales of the compounds were taken from DSC scans (10 °C min$^{-1}$). Monotropic phases are shown in parentheses.; [a] has been reported previously, [1,6] and is presented here for comparison.

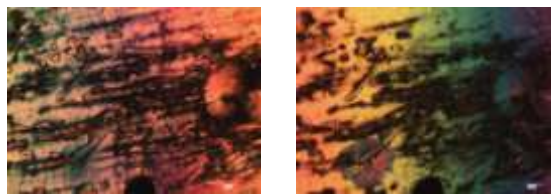

**Figure 1:** Micrographs of compound **7** at 101 °C (left) and 85 °C (right) under crossed polarizers and on untreated slides.

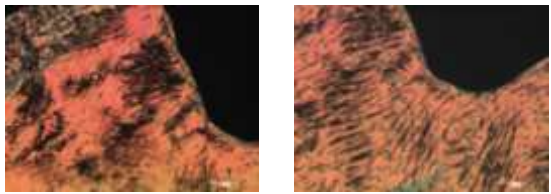

**Figure 2:** Example of textures observed $N_x$ phase on untreated slides of compound **2** in the $N_x$ phase at 33°C and 31°C.

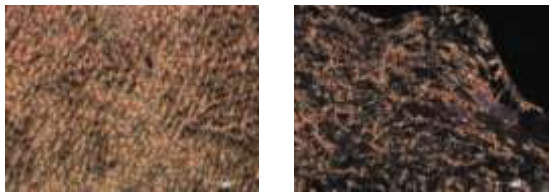

**Figure 3:** Polygonal fish scale texture of $N_x$ (a) and upon shearing (b) on untreated slides of compound **2** at 33 ˚C under crossed polarizers. Magnification x100, scale bar 100 μm.

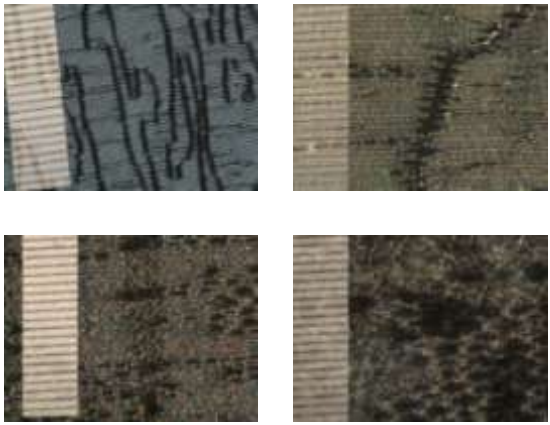

**Figure 4:** Micrographs of **CBC9CB** in planar aligned cells in the N$_x$ phase of: a) 88 ˚C, 2μm, b) 103 ˚C, 6μm c)103 ˚C, 10μ d) 104,˚ C, 25μm thickness under crossed polarisers. Indentations represent 10μm; magnification x500; crossed polarizers.

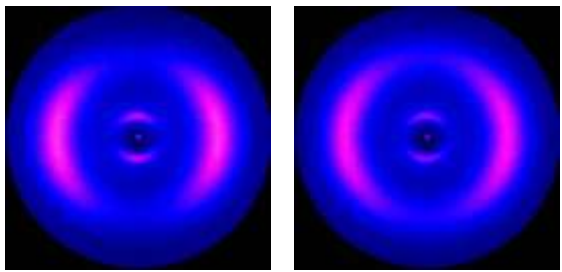

**Figure 5 :** XRD diffraction data for compound **4** at 58 °C (left) and 52 °C (right), The magnetic field is vertical. The loss of macroscopic ordering going to the $N_x$ phase is easily visually detectable, the wide angle arcs are less centred on the equator, but more spread out.

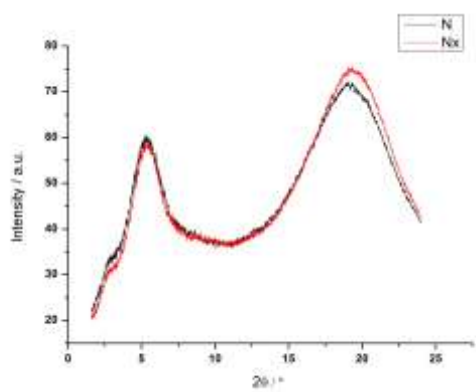

**Figure 6:** Plot of 2 theta against intensity for compound **4** in both nematic phases, recorded at 58 ˚C and 52˚C.

| Table 2 | | | | | | | | |
|---|---|---|---|---|---|---|---|---|
| Comp. | Phase | T / °C | d-spacing SAXS / Å | | | | Longitudinal Correlation Length (x) / Å | d-spacing WAXS / Å | $I_{SAX}/I_{WAX}$ |
| **1** | N | 95.0 | | | | | - | 4.7 | - |
|  | $N_x$ | 85.0 | | | | | - | 4.6 | - |
| **2** | N | 40.0 | 17.5 | | | | 12 | 4.6 | 0.69 |
|  | $N_x$ | 35.0 | 15.9 | | | | 12 | 4.6 | 0.65 |
| **3** | N | 84.0 | 18.8 | | | | 12 | 4.8 | 0.72 |
| **4** | N | 56.0 | 17.6 | | | | 13 | 4.7 | 0.86 |
|  | $N_x$ | 42.0 | 17.4 | | | | 12 | 4.6 | 0.80 |
| **5** | N | 90.0 | 21.3 | | | | 14 | 4.8 | 1.19 |
|  | $N_X$ | 80.0 | 20.8 | | | | 13 | 4.8 | 1.02 |
| **6** | N | 70.0 | 18.4 | | | | - | 4.6 | 0.90 |
|  | SmA | 50.0 | 47.4 | 22.9 | | | 1176 | 4.6 | 19.70 |
|  | SmX | 30.0 | 44.2 | 28.4 | 21.4 | 16.8 | 1166 | 4.5 | 18.13 |
| **7** | N | 99.0 | 19.2 | | | | 10 | 4.8 | 0.87 |
|  | $N_x$ | 83.0 | 19.2 | | | | 11 | 4.7 | 0.82 |
| **8** | N | 90.0 | 17.5 | | | | 14 | 4.7 | 1.04 |
|  | $N_x$ | 70.0 | 17.6 | | | | 16 | 4.6 | 0.95 |
|  | SmX | 60.0 | 18.7 | | | | - | 4.5 | - |
| **5CB** | N | 22.0 | 23.6 | 13.0 | | | 20 | 4.5 | 0.92 |

Table 2. Summary of the X-ray data; Comp. = compound.

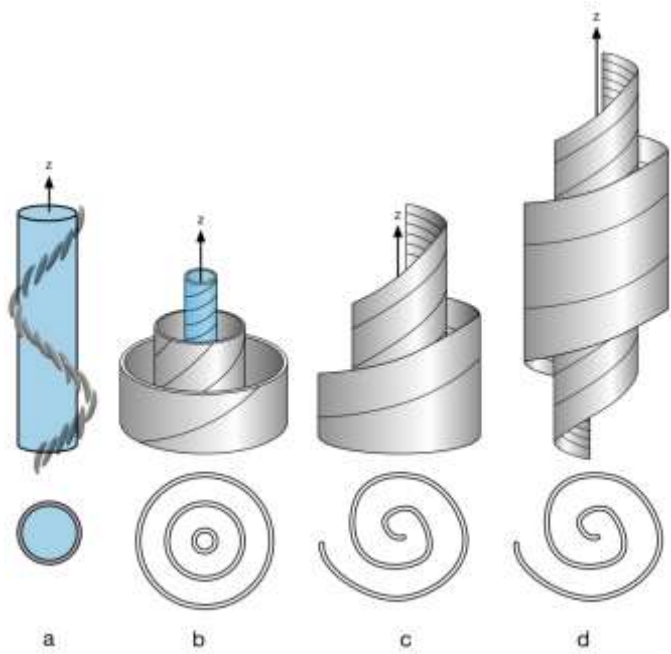

**Figure 7:** Possible arrangement of molecules. a) schematic side view of the $N_{tb}$ model, below top view; b) side view coaxial helices, below side view; c) spiralling helicoid; below top view; d) side view: spiralling helicoid with finite extension in of the z-axis, below top view.